\documentclass[a4paper,12pt]{article}
\usepackage{amsmath}
\usepackage{latexsym}
\begin{document}
\begin{titlepage}
\begin{flushright}
\textbf{UA/NPPS-02-2001}
\end{flushright}

\begin{centering}
\vspace*{1.cm}

{\large{\bf Conditional Symmetries and Phase Space Reduction}}\\
{\large{\bf towards G.C.T. Invariant Wave functions, }}\\
{\large{\bf for the Class A Bianchi Type VI \& VII Vacuum
Cosmologies}}

\vspace*{1.cm} {\bf T. Christodoulakis$^{\dag}$, G.
Kofinas$^{\S}$, G.O. Papadopoulos$^{\ddag}$}

\vspace{.7cm} {\it University of Athens, Physics Department
Nuclear \& Particle Physics Section\\
GR--15771  Athens, Greece}\\
\end{centering}

\vspace{2.cm} \abstract{The quantization of Class A Bianchi Type
VI and VII geometries --with all six scale factors, as well as the
lapse function and the shift vector present-- is considered. A
first reduction of the initial 6--dimensional configuration space
is achieved by the usage of the information furnished by the
quantum form of the linear constraints. Further reduction of the
space in which the wave function --obeying the Wheeler-DeWitt
equation-- lives, is accomplished by revealing a classical
integral of motion, tantamount to an extra symmetry of the
corresponding classical Hamiltonian. This symmetry generator
--member of a larger group-- is linear in momenta and corresponds
to G.C.T.s through the action of the automorphism group
--especially through the action of the outer automorphism
subgroup. Thus, a G.C.T. invariant wave function is found, which
depends on one combination of the two curvature invariants --which
uniquely and irreducibly, characterize the hypersurfaces
$t=const$.}

\vspace{2.2cm} \noindent
\rule[0.cm]{11.6cm}{.009cm} \\
\vspace{.3cm} e-mail: \, $\dag$ tchris@cc.uoa.gr, $\S$
gkofin@phys.uoa.gr, $\ddag$ gpapado@cc.uoa.gr
\end{titlepage}
\baselineskip=19pt

\newpage

\section{}
As it is well known, the quantum cosmology approximation consists
in freezing out all but a finite number of degrees of freedom of
the gravitational field and quantize the rest. This is done by
imposing spatial homogeneity. Thus, our --in principle-- dynamical
variables are the scale factors $\gamma_{\alpha\beta}(t)$, the
lapse function $N(t)$ and the shift vector $N^{a}(t)$, of some
spatially homogeneous geometry.

The general Class A Bianchi Type VI and VII cosmologies can be
treated on the same footing, because of their similar structure
constants.

In this sort communication, we present a complete reduction of the
initial configuration space for the above mentioned Bianchi Types;
one combination of the two curvature invariants of the
corresponding 3-space, is thus revealed as the only true degree of
freedom on which the wave function depends.

\section{}
In this work, we will quantize the known action corresponding to
the most general Class A Bianchi Type VI and VII cosmologies, i.e.
the action giving Einstein's field equations derived from the line
element
\begin{equation}
ds^{2}=(-N^{2}(t)+N_{\alpha}(t)N^{\alpha}(t))dt^{2}+2N_{\alpha}(t)\sigma^{\alpha}_{i}(x)dx^{i}dt+
\gamma_{\alpha\beta}(t)\sigma^{\alpha}_{i}(x)\sigma^{\beta}_{j}(x)dx^{i}dx^{j},
\end{equation}
where $\sigma^{\alpha}_{i}$ are the invariant basis one-forms of
the homogeneous surfaces of simultaneity $\Sigma_{t}$, satisfying
\begin{equation}
d\sigma^{\alpha}=C^{\alpha}_{\beta\gamma}~\sigma^{\beta}\wedge
\sigma^{\gamma}\Leftrightarrow \sigma^{\alpha}_{i
,~j}-\sigma^{\alpha}_{j,~i}=2C^{\alpha}_{\beta\gamma}~
\sigma^{\gamma}_{i}~\sigma^{\beta}_{j},
\end{equation}
with $C^{\alpha}_{\beta\gamma}$ being the structure constants of
the corresponding isometry group.\\ In 3 dimensions, the tensor
$C^{\alpha}_{\beta\gamma}$ admits a unique decomposition in terms
of a contravariant symmetric tensor density of weight -1,
$m^{\alpha\beta}$, and a covariant vector
$\nu_{\alpha}=\frac{1}{2}C^{\rho}_{\alpha\rho}$ as follows:
\begin{equation}
C^{\alpha}_{\beta\gamma}=m^{\alpha\delta}\varepsilon_{\delta\beta\gamma}+\nu_{\beta}\delta^{\alpha}_{\gamma}-\nu_{\gamma}\delta^{\alpha}_{\beta}.
\end{equation}
For the Class A $(\nu_{\alpha}=0)$ Bianchi Type VI
$(\varepsilon=1)$ and VII $(\varepsilon=-1)$, this matrix is
\begin{equation}
m^{\alpha\beta}=\left(\begin{array}{ccc}
  \varepsilon & 0 & 0 \\
  0 & -1 & 0 \\
  0 & 0 & 0
\end{array}\right),
\end{equation}
resulting in the following non vanishing structure constants
(\cite{1}):
\begin{equation}
\begin{array}{cc}
  C^{1}_{23}=\varepsilon & ~~~C^{2}_{13}=1.
\end{array}
\end{equation}

As is well known \cite{2}, the Hamiltonian of the above system is
$H=\widetilde{N}(t)H_{0}+N^{\alpha}(t)H_{\alpha}$, where
\begin{equation}
H_{0}=\frac{1}{2}L_{\alpha\beta\mu\nu}\pi^{\alpha\beta}\pi^{\mu\nu}+\gamma
R
\end{equation}
is the quadratic constraint, with
\begin{equation}
\begin{array}{ll}
  L_{\alpha\beta\mu\nu}=\gamma_{\alpha\mu}\gamma_{\beta\nu}+\gamma_{\alpha\nu}\gamma_{\beta\mu}-
\gamma_{\alpha\beta}\gamma_{\mu\nu} \\
  R=C^{\beta}_{\lambda\mu}C^{\alpha}_{\theta\tau}\gamma_{\alpha\beta}\gamma^{\theta\lambda}
\gamma^{\tau\mu}+2C^{\alpha}_{\beta\delta}C^{\delta}_{\nu\alpha}\gamma^{\beta\nu}+
4C^{\mu}_{\mu\nu}C^{\beta}_{\beta\lambda}\gamma^{\nu\lambda}
\end{array}
\end{equation}
(the last term of $R$ vanishes in Class A Models), $\gamma$ being
the determinant of $\gamma_{\alpha\beta}$, and
\begin{equation}
H_{\alpha}=C^{\mu}_{\alpha\rho}\gamma_{\beta\mu}\pi^{\beta\rho}
\end{equation}
are the linear constraints. Note that $\widetilde{N}$ appearing in
the Hamiltonian, is to be identified with $N/\sqrt{\gamma}$. The
canonical equations of motion, following from (6), are equivalent
to Einstein's equations derived from line element (1) (see
\cite{2}).

The quantities $H_{0}$, $H_{\alpha}$ are weakly vanishing
\cite{3}, i.e. $H_{0}\approx 0$, $H_{\alpha}\approx 0$. For all
Class A Bianchi Types ($C^{\alpha}_{\alpha\beta}=0$), it can be
seen --using the basic PBR $\{\gamma_{\alpha\beta},
\pi^{\mu\nu}\}=\delta^{\mu\nu}_{\alpha\beta}$-- that these
constraints are first class, obeying the following algebra
\begin{equation}
\begin{array}{l}
  \{H_{0}, H_{\alpha}\}=0 \\
  \{H_{\alpha},
  H_{\beta}\}=-\frac{1}{2}C^{\gamma}_{\alpha\beta}H_{\gamma},
\end{array}
\end{equation}
which ensures their preservation in time, i.e. $\dot{H}_{0}\approx
0$, $\dot{H}_{\alpha}\approx 0$, and establishes the consistency
of the action.

If we follow Dirac's general proposal \cite{3} for quantizing this
action, we have to turn $H_{0}$, $H_{\alpha}$, into operators
annihilating the wave function $\Psi$.

In the Schr\"{o}dinger representation
\begin{equation}
\begin{array}{l}
  \gamma_{\alpha\beta}\rightarrow
\widehat{\gamma}_{\alpha\beta}=\gamma_{\alpha\beta} \\
  \pi^{\alpha\beta}\rightarrow
\widehat{\pi}^{\alpha\beta}=-i\frac{\partial}{\partial\gamma_{\alpha\beta}},
\end{array}
\end{equation}
with the relevant operators satisfying the basic Canonical
Commutation Relations (CCR) --corresponding to the classical
ones--:
\begin{equation}
[\widehat{\gamma}_{\alpha\beta},
\widehat{\pi}^{\mu\nu}]=i\delta^{\mu\nu}_{\alpha\beta}=\frac{i}{2}
(\delta^{\mu}_{\alpha}\delta^{\nu}_{\beta}+\delta^{\mu}_{\beta}\delta^{\nu}_{\alpha}).
\end{equation}

In general, via the method of characteristics \cite{4}, the
quantum version of the three independent linear constraints can be
used to reduce the dimension of the initial configuration space
from 6 ($\gamma_{\alpha\beta}$) to 3 (combinations of
$\gamma_{\alpha\beta}$), i.e. $\Psi=\Psi(q^{1},q^{2},q^{3})$
\cite{5}, where
\begin{equation}
\begin{array}{l}
  q^{1}=C^{\alpha}_{\mu\kappa}C^{\beta}_{\nu\lambda}\gamma^{\mu\nu}\gamma^{\kappa\lambda}\gamma_{\alpha\beta} \\
  q^{2}=C^{\alpha}_{\beta\kappa}C^{\beta}_{\alpha\lambda}\gamma^{\kappa\lambda} \\
  q^{3}=\gamma.
\end{array}
\end{equation}
One can prove \cite{6,7} that the only G.C.T. (gauge) invariant
quantities, which uniquely and irreducibly, characterize a
3-dimensional geometry admitting the Class A Type VI and VII
symmetry groups, are the quantities $q^{1}$ and $q^{2}$.
An outline of the proof, is as follows:\\
Let two hexads $\gamma^{(1)}_{\alpha\beta}$ and
$\gamma^{(2)}_{\alpha\beta}$ be given, such that their
corresponding $q^{1},~q^{2}$ are equal. Then \cite{6}, there
exists an automorphism matrix $\Lambda$ (i.e. a matrix satisfying
$C^{\alpha}_{\mu\nu}\Lambda^{\kappa}_{\alpha}=C^{\kappa}_{\rho\sigma}\Lambda^{\rho}_{\mu}\Lambda^{\sigma}_{\nu}$)
connecting them, i.e.
$\gamma^{(1)}_{\alpha\beta}=\Lambda^{\mu}_{\alpha}\gamma^{(2)}_{\mu\nu}\Lambda^{\nu}_{\beta}$
(the inverse is easily seen to be true, as well). But, as it has
been shown in the appendix of \cite{7}, this kind of changes on
$\gamma_{\alpha\beta}$, can be seen to be induced by spatial
diffeomorphisms. Thus, 3-dimensional Class A Type VI and VII
geometries, are uniquely and irreducibly characterized by some
values of these two $q$~'s.

In terms of the three $q$~'s, one can define the following
--according to \cite{5}-- induced ``physical'' metric, given by
the relation
\begin{equation}
g^{ij}=L_{\alpha\beta\mu\nu}\frac{\partial q^{i} }{\partial
\gamma_{\alpha\beta}}\frac{\partial q^{j}}{\partial
\gamma_{\mu\nu}}=\left(\begin{array}{ccc}
  5q^{1}q^{1}-16q^{2}q^{2} & q^{1}q^{2} & q^{1}q^{3} \\
  q^{1}q^{2} & q^{2}q^{2} & q^{2}q^{3} \\
  q^{1}q^{3} & q^{2}q^{3} & -3q^{3}q^{3}
\end{array}\right).
\end{equation}
Note that the first-class algebra satisfied by $H_{0}$,
$H_{\alpha}$, ensures that indeed, all components of $g^{ij}$ are
functions of the $q^{i}$.

According to K\v{u}char's and Hajicek's \cite{5} prescription, the
``kinetic'' part of $H_{0}$ is to be realized as the conformal
Laplacian, based on the ``physical'' metric (13). However, in the
case under discussion here, there is a further reduction.

For full pure gravity, K\v{u}char \cite{8} has shown that there
are no other first-class functions, homogeneous and linear in the
momenta, except the linear constraints. If however, we impose
extra symmetries (e.g. the Class A Type VI and VII --here
considered), such quantities may emerge --as it will be shown. We
are therefore --according to Dirac \cite{3}-- justified to seek
the generators of these extra symmetries, whose quantum-operator
form will be imposed as additional conditions on the wave
function. Thus, these symmetries are expected to lead us to the
final reduction, by revealing the true degrees of freedom. Such
quantities are, generally, called in the literature ``Conditional
Symmetries'' \cite{8}.

The automorphism group for the Class A Type VI and VII, is
described by the following 4 generators --in matrix notation and
collective form:
\begin{equation}
\lambda^{\alpha}_{(I)\beta}=\left(
\begin{array}{ccc}
  a  & \varepsilon~c & b \\
  c & a & d \\
  0 & 0 & 0
\end{array}\right),
\end{equation}
with the defining property
\begin{equation}
C^{\alpha}_{\mu\nu}\lambda^{\kappa}_{\alpha}=C^{\kappa}_{\mu\sigma}\lambda^{\sigma}_{\nu}+C^{\kappa}_{\sigma\nu}\lambda^{\sigma}_{\mu}.
\end{equation}
From these matrices, we can construct the linear --in momenta--
quantities
\begin{equation}
A_{(I)}=\lambda^{\alpha}_{(I)\beta}\gamma_{\alpha\rho}\pi^{\rho\beta}.
\end{equation}
Three of these are the $H_{\alpha}$'s, since
$C^{\alpha}_{(\rho)\beta}$ correspond to the inner automorphism
subgroup --designated by the $c$, $b$ and $d$ parameters in
$\lambda^{\alpha}_{(I)\beta}$. The remaining is the generator of
the outer automorphisms and is given by the essentially unique
matrix
\begin{equation}
\epsilon^{\alpha}_{\beta}=\left(\begin{array}{ccc}
  a & 0 & 0 \\
  0 & a & 0 \\
  0 & 0 & 0
\end{array}\right).
\end{equation}
The corresponding --linear in momenta-- quantity is
\begin{equation}
E=\epsilon^{\alpha}_{\beta}\gamma_{\alpha\rho}\pi^{\rho\beta}.
\end{equation}
It is straightforward to calculate the Poisson Brackets of $E$
with $H_{0},~H_{\alpha}$:
\begin{equation}
\begin{array}{l}
  \{E, H_{\alpha}\}=-\frac{1}{2}\lambda^{\beta}_{\alpha}H_{\beta} \\
  \{E, H_{0}\}=-2a\gamma R=-2a\gamma (q^{1}+2q^{2}).
\end{array}
\end{equation}

At this point, it is crucial to observe that we can construct a
classical integral of motion, i.e. an extra gauge symmetry of the
corresponding classical action: notice that the trace of the
canonical momenta, $\gamma_{\mu\nu}\pi^{\mu\nu}$, has vanishing PB
with $H_{\alpha},~E$, and a --similar to $E$-- PB with $H_{0}$
equal to $2\gamma R$; thus, if we define
\begin{equation}
T=E-a\gamma_{\mu\nu}\pi^{\mu\nu},
\end{equation}
we can easily derive the following PB of $T$ with
$H_{0},~H_{\alpha}$:
\begin{equation}
\begin{array}{l}
  \{T, H_{0}\}=0 \\
  \{T, H_{1}\}=-\frac{a}{2}H_{1}\approx 0 \\
  \{T, H_{2}\}=-\frac{a}{2}H_{2}\approx 0\\
  \{T, H_{3}\}=0.
\end{array}
\end{equation}

The quantity $T$, is thus revealed to be first-class, and
therefore an integral of motion (since the Hamiltonian, is a
linear combination of the constraints):
\begin{equation}
\dot{T}=\{T,H\}\approx 0 \Rightarrow T=\text{const}=C_{T}.
\end{equation}
The quantum version of $T$, is taken to be \cite{5}:
\begin{equation}
\widehat{T}=\epsilon^{\alpha}_{\beta}\gamma_{\alpha\rho}\frac{\partial}{\partial
\gamma_{\beta\rho}}-a\gamma_{\alpha\beta}\frac{\partial}{\partial
\gamma_{\alpha\beta}}
\end{equation}
(without any loss of generality --see (17)-- we can safely
suppress $a$, whenever convenient, by setting it to 1).

Following the spirit of Dirac, we require:
\begin{equation}
\widehat{T}\Psi=\epsilon^{\alpha}_{\beta}\gamma_{\alpha\rho}\frac{\partial
\Psi}{\partial
\gamma_{\beta\rho}}-a\gamma_{\alpha\beta}\frac{\partial
\Psi}{\partial
\gamma_{\alpha\beta}}=(q^{1}\frac{\partial\Psi}{\partial
q^{1}}+q^{2}\frac{\partial\Psi}{\partial
q^{2}}-\gamma\frac{\partial\Psi}{\partial \gamma})=C_{T}\Psi.
\end{equation}
The general solution \cite{4} to the above equation has the form
\begin{equation}
\Psi=\gamma^{-C_{T}}\Theta(\frac{q^{1}}{q^{2}},\gamma q^{2}),
\end{equation}
$\Theta$ being an arbitrary function in its arguments.

Now, the number of our dynamical variables, is reduced from 3
($q^{i}, i=1,2,3$) to 2, namely the combinations
$w^{1}=q^{1}/q^{2}$ and $w^{2}=q^{2}q^{3}$. So, we have a further
reduction of the 3-dimensional configuration space spanned by the
3 $q$~'s. Again, in terms of the $w$~'s, the finally reduced
``physical'' --although singular-- metric, is given by the
following relation
\begin{equation}
s^{kl}=g^{ij}\frac{\partial w^{k}}{\partial q^{i}}\frac{\partial
w^{l}}{\partial q^{j}}=\left(\begin{array}{cc}
  -16+4(w^{1})^{2} & 0 \\
  0 & 0
\end{array}\right).
\end{equation}
The singular character of this metric is not unexpected; its
origin lies in the fact that
$L^{\alpha\beta\gamma\delta}T_{\alpha\beta}T_{\gamma\delta}=0$,
where
$L^{\alpha\beta\gamma\delta}=\frac{1}{4}(\gamma^{\alpha\gamma}\gamma^{\beta\delta}+
\gamma^{\alpha\delta}\gamma^{\beta\gamma}-2\gamma^{\alpha\beta}\gamma^{\gamma\delta})$
is the covariant supermetric (inverse to
$L_{\alpha\beta\gamma\delta}$) and $T_{\alpha\beta}$ are the
components of $T$, seen as vector field in the initial superspace
spanned by $\gamma_{\alpha\beta}$. Indeed, it is known that
reducing to null surfaces entails all sort of peculiarities.

So far, the degrees of freedom are two ($w^{1},~w^{2}$). The
vanishing of the $s^{22}$ component indicates that $w^{2}$ is not
dynamical at the quantum level. This fact has its analogue at the
classical level; indeed, consider the quantity:
\begin{equation}
\Omega=\frac{(\gamma q^{2})^{\cdot}}{\gamma
q^{2}}=L_{\alpha\beta\mu\nu}\gamma^{\mu\nu}\pi^{\alpha\beta}-L_{\kappa\lambda\rho\sigma}
\pi^{\rho\sigma}\frac{C^{\alpha}_{\beta\mu}C^{\beta}_{\alpha\nu}\gamma^{\mu\kappa}
\gamma^{\nu\lambda}}{q^{2}}.
\end{equation}
In (27), the transition from velocity phase space to momentum
phase space has been made using the Hamiltonian (6). It is
straightforward to verify that:
\begin{equation}
\Omega=\frac{2}{a}T-\frac{4\varepsilon
\gamma^{23}}{q^{2}}H_{1}-\frac{4\varepsilon
\gamma^{13}}{q^{2}}H_{2}.
\end{equation}
Taking into account the weak vanishing of the linear constraints,
it is deduced that
\begin{equation}
\Omega=\frac{2}{a}T=\frac{2C_{T}}{a}=2C_{T},
\end{equation}
if we set $a=1$.\\
Another way to show that $\Omega$ is constant, is the following:
\begin{equation}
\{\Omega, H\}=\{\frac{\{\gamma q^{2}, H\}}{\gamma q^{2}},
H\}=\frac{4\gamma}{(\gamma
q^{2})^{2}}(4\gamma_{11}H_{1}^{2}+4\gamma_{22}H_{2}^{2}-8\varepsilon\gamma_{12}H_{1}H_{2}).
\end{equation}
Again, the weak vanishing of the linear constraints, ensures that
$\Omega$ is constant.

Using (27), (29) and the action, we have:
\begin{equation}
\gamma q^{2}=C_{1}exp\{2C_{T}\int\widetilde{N}(t)dt\}.
\end{equation}
Now, returning to the quantum domain, we observe that out of the
three arguments of the wave function given in (25), only
$q^{1}/q^{2}$ is G.C.T. invariant --in the sense previously
explained. This suggests that we must somehow eliminate $\gamma,
~\gamma q^{2}$. To this end, we adopt the value zero for the
classical constant $C_{T}$. This amounts to restricting to a
2-parameter subspace of the classical space of solutions, spanned
by the 3 essential constants ~\cite{9}. This means that we base
our quantum theory on this subspace and decree the wave function,
to be applicable to all configurations (classical or not). The
benefit of such an action, is twofold: $\gamma^{-C_{T}}$ drops
out, while at the same time $w^{2}\equiv\gamma q^{2}$ is set equal
to the constant $C_{1}$ --see (31). These facts, along with the
obligation that no derivatives with respect to $w^{2}$, are to
enter the Wheeler-DeWitt equation --see (26)--, allow us to arrive
at the following form for the wave function (25):
\begin{equation}
\Psi=\Theta(\frac{q^{1}}{q^{2}},C_{1}),
\end{equation}
and of course
\begin{equation}
w^{2}=q^{2}q^{3}=\text{const}=C_{1}.
\end{equation}
Now, the final reduction of the configuration space is achieved.
Our dynamic variable is the ratio $q^{1}/q^{2}$, which is a
combination of the only curvature invariants existing in Class A
Bianchi Type VI and VII, and emerges as the only true quantum
degree of freedom.

Consequently, following the spirit of \cite{5}, we have to
construct the quantum analogue of $H_{0}$ as the conformal
Laplacian, based on the non-singular part of the ``physical''
metric (26), i.e.:
\begin{equation}
\widehat{H}_{0}=-\frac{1}{2}\nabla^{2}_{c}+w^{2}(2+w^{1}),
\end{equation}
where
\begin{equation}
\nabla^{2}_{c}=\nabla^{2}=\frac{1}{\sqrt{s_{11}}}~\partial_{w^{1}}
\{\sqrt{s_{11}}~s^{11}~\partial_{w^{1}}\}
\end{equation}
is the 1-dimensional Laplacian based on $s_{11}$
($s^{11}s_{11}=1$). Note that in 1-dimension the conformal group
is totally contained in the G.C.T. group, in the sense that any
conformal transformation of the metric can not produce any change
in the --trivial-- geometry and is thus reachable by some G.C.T.
Therefore, no extra term in needed in (35), as it can also
formally be seen by taking the limit $d=1,~R=0$ in the general
definition:
\begin{displaymath}
\nabla^{2}_{c}\equiv\nabla^{2}+\frac{(d-2)}{4(d-1)}R=\nabla^{2}.
\end{displaymath}

Thus, the Wheeler-DeWitt equation, reads:
\begin{equation}
\widehat{H}_{0}\Psi=\sqrt{(w^{2}-4)}~
\partial_{w}\{\sqrt{(w^{2}-4)}~\partial_{w}\}\Psi-\frac{C_{1}(2+w)}{2}\Psi=0,
\end{equation}
where, for simplicity, we set $w^{1}\equiv w$ and $w^{2}=C_{1}$.\\
Using the transformation $w=2cosh(z)$, the previous equation takes
the form
\begin{equation}
\frac{\partial^{2}\Psi}{\partial z^{2}}-(C_{1}+cosh(z))\Psi=0.
\end{equation}
The solutions to this family of equations are the Mathieu Modified
Functions --see \cite{10} and references therein, for an extended
treatise, in various cases.

As for the measure, it is commonly accepted that there is not a
unique solution. A natural choice, is to adopt the measure that
makes the operator in (36) hermitian, that is
$\mu(w)=\frac{1}{\sqrt{w^{2}-4}}$, or in the variable $z$,
$\mu(z)=1$. However, the solutions to (37) can be seen to
violently diverge for various values of $z \in [0,\infty)$, which
is the classically allowed region. If we wish to avoid this
difficulty, we can abandon hermiticity, especially in view of the
fact that we are interested in the zero eigenvalues of the
operator, and thus does not make any harm to lose realness of the
eigenvalues. If we adopt this attitude, we can find suitable
measures, e.g. $\mu(z)=e^{-z^{2}}$. The probability density
$\rho(z)=\mu(z)|\Psi(z)|^{2}$ is now finite, enabling one to
assign a number between $0$ and $1$ to
each Class A Type VI and VII geometry.\\
Another feature of the reduction to (37) is that the final
dynamical argument of $\Psi$ is the ratio $q^{1}/q^{2}$, which is
of degree zero in the scale factors, as seen from equation (12).
This consists a kind of build-in regularization with respect to
the volume of the 3-space. Moreover, the solutions to this
equation exhibit an increasingly oscillatory behaviour, as $C_{1}$
increases. This is most welcomed and expected in view of $C_{1}$
being some kind of measure of the 3-volume, since it contains
$\gamma$ --see (33).

\section{}
In this work, we were able to express the wave function for the
Class A Bianchi Type VI and VII Vacuum cosmologies, in terms of a
unique quantum degree of freedom, i.e. the ratio of the two
curvature invariants ($q^{1}, q^{2}$), which completely and
irreducibly characterize the 3-geometries under discussion. At a
first stage, this was done by imposing the quantum versions of the
first-class linear constraints, as conditions on the wave function
$\Psi$. At a second stage, the indirect usage --through a
classical integral of motion and its quantum version-- of the
outer automorphisms reduced the dimension of the configuration
space --spanned by the $q$~'s-- from 3 to 2. The final reduction
came through the assumption/condition that one ought to end up
with a G.C.T. invariant wave function. Thus, the classically
conserved value of $T$ --which was shown to be related to the
other degree of freedom, i.e. the product $q^{2}q^{3}$-- was set
to zero. Anyway, this is not such a blunder, since the ultimate
goal of finding a $\Psi$, is to give weight to all states --being
classical ones, or not. So, in conclusion, we see that not only
the true degree of freedom is isolated, but also the time problem
has been solved --in the sense that a square integrable wave
function $\Psi$ is found. This is accomplished by revealing all
the --hidden-- gauge symmetries of the system. This wave function
is G.C.T. invariant and well defined on any spatially homogeneous
3-geometry.

A similar situation holds for Class A Type II; objects analogous
to $T$ also exist, and thus, a reduction to the variable $\gamma
q^{1}$ is possible, with the help of this $T$ plus another one
--the $\Omega$-- (see \cite{11}) and the 2 independent
$H_{\alpha}$'s ($q^{1}=q$ is the unique independent curvature
invariant for these homogeneous 3-geometries). The situation
concerning Type VIII and IX is more difficult: there are no outer
automorphisms, and consequently, no object analogous to $T$,
exists. The $H_{\alpha}$'s suffice to reduce the configuration
space to $q^{1}$, $q^{2}$ and $q^{3}=Det[m]/\sqrt{\gamma}$, now
needed to specify the 3-geometry. The reduced supermetric is still
Lorentzian, leading to a hyperbolic Wheeler-DeWitt equation. This
fact is generally considered as a drawback, since it prohibits the
ensuing wave function from being square integrable.

A very interesting case is that of Bianchi Type I; there, the
automorphism group is the whole $GL(3,~\Re)$. This peculiarity may
prove extremely helpful in reducing the initial configuration
space through the many integrals of motion. A work of ours on this
issue, is in progress and it will be available as soon as
possible.

\vspace{1cm} \noindent \textbf{Acknowledgments}\\ The authors
acknowledge financial support by the University of Athens' Special
Account for the Research.

\end{document}